\documentclass{mn2e}
\usepackage{psfig}

\title[The 2dFGRS: Voids and hierarchical scaling models]
{The 2dF Galaxy Redshift Survey: Voids and hierarchical scaling models}

\author[Croton et al.]{
\parbox[t]{\textwidth}{
Darren J. Croton$^{1,2}$,
Matthew Colless$^{2,3}$,
Enrique Gazta\~{n}aga$^{4,5}$,
Carlton M. Baugh$^6$,
Peder Norberg$^7$,
I. K.\ Baldry$^8$,
J. Bland-Hawthorn$^3$,
T. Bridges$^9$, 
R. Cannon$^3$, 
S. Cole$^6$, 
C. Collins$^{10}$, 
W. Couch$^{11}$, 
G. Dalton$^{12,13}$,
R. De Propris$^2$,
S. P.\ Driver$^2$, 
G. Efstathiou$^{14}$, 
R. S.\ Ellis$^{15}$, 
C. S.\ Frenk$^6$, 
K. Glazebrook$^8$, 
C. Jackson$^{16}$,
O. Lahav$^{14,17}$, 
I. Lewis$^{12}$, 
S. Lumsden$^{18}$, 
S. Maddox$^{19}$,
D. Madgwick$^{20}$,
J. A.\ Peacock$^{21}$,
B. A.\ Peterson$^2$, 
W. Sutherland$^{21}$,
K. Taylor$^{15}$
(The 2dFGRS Team)
}
\vspace*{6pt} \\ 
$^1$Max-Planck-Institut f\"ur Astrophysik, D-85740 Garching, Germany \\
$^2$Research School of Astronomy \& Astrophysics, The Australian 
    National University, Weston Creek, ACT 2611, Australia \\
$^3$Anglo-Australian Observatory, P.O.\ Box 296, Epping, NSW 2111,
    Australia\\  
$^4$INAOE, Astrofisica, Tonantzintla, Apdo Postal 216 y 51, Puebla 7200,
    Mexico \\
$^5$Institut d'Estudis Espacials de Catalunya, ICE/CSIC, Edf.
    Nexus-104-c/Gran Capita 2-4, 08034 Barcelona, Spain  \\
$^6$Department of Physics, University of Durham, South Road, 
    Durham DH1 3LE, UK \\ 
$^7$ETHZ Institut f\"ur Astronomie, HPF G3.1, ETH H\"onggerberg, CH-8093
       Z\"urich, Switzerland \\
$^8$Department of Physics \& Astronomy, Johns Hopkins University,
       Baltimore, MD 21118-2686, USA \\
$^{9}$Department of Physics, Queen's University, Kingston, 
    Ontario K7L 3N6, Canada \\
$^{10}$Astrophysics Research Institute, Liverpool John Moores University,  
    Twelve Quays House, Birkenhead, L14 1LD, UK \\
$^{11}$Department of Astrophysics, University of New South Wales, Sydney, 
    NSW 2052, Australia \\
$^{12}$Department of Physics, University of Oxford, Keble Road, 
    Oxford OX1 3RH, UK \\
$^{13}$Space Science \& Technology Division, Rutherford Appleton Laboratory, 
    Chilton OX11 0QX, UK \\
$^{14}$Institute of Astronomy, University of Cambridge, Madingley Road,
    Cambridge CB3 0HA, UK \\
$^{15}$Department of Astronomy, California Institute of Technology, 
    Pasadena, CA 91025, USA \\
$^{16}$CSIRO Australia Telescope National Facility, PO
    Box 76, Epping, NSW 1710, Australia \\
$^{17}$Department of Physics and Astronomy, University College London, 
    Gower Street, London WC1E 6BT, UK \\
$^{18}$Department of Physics, University of Leeds, Woodhouse Lane,
       Leeds, LS2 9JT, UK \\
$^{19}$School of Physics \& Astronomy, University of Nottingham,
       Nottingham NG7 2RD, UK \\
$^{20}$Department of Astronomy, University of California, Berkeley, 
       CA 94720, USA \\
$^{21}$Institute for Astronomy, University of Edinburgh, Royal Observatory, 
       Blackford Hill, Edinburgh EH9 3HJ, UK.
\vspace{-0.5cm}
}

\date{Accepted ---. Received ---;in original form ---}

\pagerange{\pageref{firstpage}--\pageref{lastpage}}
\pubyear{2004}

\def\Nx{\bar{N} \bar{\xi}}
\def\Nbar{\bar{N}}
\def\xibar{\bar{\xi}}
\newcommand{\nc}{\newcommand}

\nc{\be}[1]{\begin{equation}\mbox{$\label{#1}$}}
\nc{\bea}[1]{\begin{eqnarray} \mbox{$\label{#1}$}}
\nc{\Section}[2]{\section{#2}\label{#1}}
\nc{\Bibitem}[1]{\bibitem{#1}}
\nc{\Label}[1]{\label{#1}}

\nc{\vev}[1]{\langle #1 \rangle}

\nc{\eea}{\end{eqnarray}}
\nc{\ee}{\end{equation}}
\nc{\eeq}{\end{equation}}

\newcommand{\plotone}[1]
           {\centering \leavevmode \psfig{file=#1,width=\columnwidth,clip=}}

\newcommand{\plotfull}[1]
           {\centering \leavevmode \psfig{file=#1,width=\textwidth,clip=}}

\begin{document}

\maketitle

\label{firstpage}

\begin{abstract}

We measure the redshift space reduced void probability function (VPF)
for 2dFGRS volume limited galaxy samples covering the absolute
magnitude range \smash{$M_{b_{\rm J}}-5\log_{10}h=-18$} to $-22$.
Theoretically, the VPF connects the distribution of voids to the
moments of galaxy clustering of all orders, and can be used to
discriminate clustering models in the weakly non-linear regime.  The
reduced VPF measured from the 2dFGRS is in excellent agreement with
the paradigm of hierarchical scaling of the galaxy clustering moments.
The accuracy of our measurement is such that we can rule out, at a
very high significance, popular models for galaxy clustering, including
the lognormal distribution. We demonstrate that the negative binomial
model gives a very good approximation to the 2dFGRS data over a wide
range of scales, out to at least \smash{$20h^{-1}$Mpc}.  Conversely,
the reduced VPF for dark matter in a $\Lambda$CDM universe does appear
to be lognormal on small scales but deviates significantly beyond
\smash{$\sim 4h^{-1}$Mpc}.  We find little dependence of the 2dFGRS
reduced VPF on galaxy luminosity.  Our results hold independently in
both the north and south Galactic pole survey regions.

\end{abstract}

\begin{keywords}
galaxies: statistics, clustering; cosmology: theory, large-scale structure, voids.
\end{keywords}

\section{Introduction}

The galaxy distribution on the largest scales display striking
geometrical features, such as walls, filaments and voids.  These
features contain a wealth of information about both the linear and
non-linear evolution of galaxy clustering.  The nature of such
clustering is dependent on many large and small scale effects, such as
the cosmological parameters, galaxy and cluster environmental effects
and history, the underlying dark matter distribution, and the way in
which the dark and luminous components of the Universe couple and
evolve.  By probing the lower and higher orders of galaxy clustering,
one thus hopes to shed light on those physical processes on which the
clustering is dependent.

The traditional tool used to analyse such distributions has been the
2-point correlation function (Davis \& Peebles 1983, Davis et
al. 1988, Fisher et al. 1994, Loveday et al. 1995, Norberg et
al. 2001, Zehavi et al. 2002), providing a description of clustering
at the lowest orders.  However despite its usefulness, the 2-point
correlation function only provides a full clustering description in
the case of a Gaussian distribution.  A more complete account of
clustering must include correlation functions of higher orders,
although these are often difficult to extract (see Croton et al. 2004
and Baugh et al. 2004 for an analysis of galaxy clustering in the
2dFGRS up to sixth order).

In light of this researchers have looked towards other clustering
statistics to glean higher-order information from a galaxy
distribution.  Historically, many astronomers have favoured using void
statistics (e.g. Fry 1986, Maurogordato \& Lachieze-Rey 1987, Balian
\& Schaeffer 1989, Fry et al. 1989, Bouchet et al. 1993, Gazta\~naga
\& Yokoyama 1993, Vogeley et al. 1994).  This approach is useful in
that results are easily obtainable and are well supported by a solid
theoretical framework (White 1979, Fry 1986, Balian \& Schaeffer
1989), which directly relates the void distribution to that of galaxy
clustering of higher orders.

In this paper we employ the completed 2dFGRS dataset to undertake a
detailed analysis of the void distribution using the reduced void
probability function.  We rely heavily on the well established
theoretical framework which connects the void distribution with galaxy
clustering of all orders (Eq.~1 below).  The distribution of voids and
the moments of galaxy clustering of all orders are known to be
intimately linked, and the study of one can reveal information about
the other which would otherwise be difficult to measure.  Our goal is
thus to use the reduced void probability function to investigate if
galaxy clustering in the 2dFGRS obeys a hierarchy of scaling, and on
what physical scales this scaling holds.  We explore a number of
phenomenological models of galaxy clustering which exhibit
hierarchical scaling, and use these models to help clarify the way in
which higher-order clustering is constructed\footnote{Recently Hoyle
et al. (2004) also measured the VPF of the 2dFGRS galaxy distribution,
however their analysis focused more on the physical properties of voids
in the 2dFGRS volume, rather than the hierarchical nature of galaxy
clustering itself.}.

This paper is organised as follows.  In Section~2 we give a brief
review of the theory behind the void statistics to be employed in our
analysis.  In Section~3 we present the 2dFGRS data set, and in
Section~4 the counts-in-cells method we use to measure the void
statistics is explained.  Our results are presented in Section~5, and
in Section~6 we provide a discussion and summary of our conclusions.
Throughout, we adopt standard present day values of the cosmological
parameters to compute comoving distance from redshift:
a density parameter $\Omega_{m}=0.3$ and a cosmological constant
$\Omega_{\Lambda}=0.7$.

\section{Void Statistics}

\subsection{The Void Probability Function}

For a given distribution of galaxies, the count probability
distribution function (CPDF), $P_N(V)$, is defined as the probability
of finding exactly $N$ galaxies in a cell of volume $V$ randomly
placed within the sample.  In the case where $N=0$ we have the void
probability function (VPF), $P_0(V)$.  A choice of spherical cells
with which to sample the distribution makes $P_0$ a function of sphere
radius $R$ only.  The VPF can be related to the hierarchy of $p$-point
correlation functions by (White 1979):
\begin{equation}
P_0(R)=\exp\Bigg[\sum_{p=1}^{\infty} \frac{(\ -\bar{N}(R)\ )^p}{p!}
\bar{\xi}_p (R) \Bigg]~.
\end{equation}
Here $\bar{N}$ is the average number of objects in a cell of volume
$V$, and $\bar{\xi}_p$ is the $p^{th}$ order correlation function
averaged over $V$.  A completely random (Poisson) distribution has
$\xi_p\equiv 0$ for all $p>1$, and thus $P_0$ reduces to a simple
analytic expression:
\begin{equation}
P_{0_P}(R)=\exp [-\bar{N}(R) ]~.
\end{equation}
Any departure from this relation is therefore a signature of the
presence of clustering.

\subsection{Hierarchical Scaling}

The idea that higher-order clustering arises in a \emph{hierarchical}
fashion from the 2-point correlation function appears naturally in
perturbation theory and also in the highly non-linear regime of
gravitational clustering (e.g. Peebles 1980), and is supported by much
observational evidence (e.g.  Maurogordato \& Lachieze-Rey 1987, Fry
et al. 1989, Gazta\~naga 1992, Bouchet et al. 1993, Bonometto et
al. 1995, Benoist et al. 1999, see Bernardeau et al. 2002 for a
review).  The concept can be generalised by assuming that each
$p$-point correlation function depends only on the product of the
2-point correlation function and a dimensionless scaling coefficient,
$S_p$:
\begin{equation}
\bar{\xi_p}(R)=S_p \bar{\xi}^{p-1} (R)~,
\end{equation}
where we have dropped the subscript $2$ for the 2-point correlation
function on the right-hand side for convenience (see Baugh et al. 2004
and Croton et al. 2004 for the measured values of $S_p$ up to $p=6$ in
redshift space for the 2dFGRS).

The hierarchical idea is directly applicable to the VPF, which is
itself dependent on an infinite sum of $p$-point correlation
functions.  The hierarchical assumption allows us to remove the
higher-order correlation functions from Eq.~1:
\begin{equation}
P_0(R)=\exp\Bigg[\sum_{p=1}^{\infty} \frac{(-\bar{N})^p}{p!}  S_p
\bar{\xi}^{p-1} \Bigg]~.
\end{equation}
Furthermore, the above scaling relation allows us to
express the VPF as a function of $\bar{N} \bar{\xi}$ only, where the
scaling variable $\bar{N} \bar{\xi}$ approximately represents the
average number of galaxies in a cell \emph{in excess} of that expected
given the mean density of the sample.  We formalise this idea by
firstly considering the analytic VPF expression for a purely random
sample (Eq.~2).  For the hierarchical situation, we can define a
parameter $\chi$ with $P_0=e^{-\bar{N} \chi}$, called the
\emph{reduced void probability function} (see Fry 1986):
\begin{equation}
\chi\ =\ -\ln(P_0)\ /\ \bar{N}~.
\end{equation}
We note here that, independent of the hierarchical assumption, $\chi$
normalises out the Poisson contribution to the distribution, and it is
clear that the effects of clustering will appear as values of
$\chi<1$.  Combining Eq.~4 and 5, the reduced VPF takes the form
\be{chiexp}
\chi(\bar{N} \bar{\xi}) = \sum_{p=1}^{\infty} \frac{S_p}{p!}
(-\bar{N} \bar{\xi})^{p-1}~.
\ee
This exhibits the scaling advertised above, and the shape of
$\chi(\bar{N} \bar{\xi})$ thus characterises the distribution of
voids.  If the scaling relation assumption holds, we expect different
galaxy samples of different density and clustering strength to all
collapse onto one universal curve, since all are a function of the
same scaling variable.  The curve will not be universal for
different magnitude ranges if it turns out that the coefficients $S_p$
are a strong function of galaxy magnitude.  The values of $S_p$ have
recently been shown to depend at best only weakly on magnitude (see
Croton et al. 2004).

In the hierarchical picture, when $\Nx \ll 1$ one always recovers the
Poisson VPF, $\chi(\Nx)=1$, regardless of the actual clustering
pattern or its strength.  In the regime where $\Nx<1$ we see from
Eq.~6 that the reduced void probability function is dominated by the
Gaussian contribution: $1-\frac{1}{2} \bar{N} \bar{\xi}$.  Thus the
interesting observational window, where we can separate different
clustering models, comes for values of $\Nx$ larger than unity.  In
practice, this only seems to happen at scales $R$ larger than a few
$h^{-1}$Mpc, where $\Nbar \sim R^{3}$ is large and dominates $\xibar
\sim R^{-2}$.  On smaller scales, where $\xibar>1$, $\Nx$ will always
be small, and galaxy samples will typically be too sparse to show
measurable deviations from the Gaussian contribution.
Thus, it should be stressed that the VPF is a good discriminant of
weakly non-linear clustering only. In the highly non-linear regime
voids do not provide us with much information.

Although the expansion given in Eq.~\ref{chiexp} is technically only
valid for small values of $\Nx$, the implications for clustering do
extend beyond this. For large values of $\Nx$ models with different
hierarchical amplitudes $S_p$ give different reduced void
probabilities $\chi$: as $\Nx$ increases the value of $\chi$ gets
smaller and the resulting VPF gets larger (with respect to the
corresponding Poisson case). The Gaussian CPDF ($S_p=0$) produces the
smallest values of $\chi$ and therefore the largest deviations in the
VPF.  As we will illustrate with the models below, larger values of
$S_p>0$ will result in larger values of $\chi(\Nx)$.

\subsection{Phenomenological Models}

\begin{figure}
\plotone{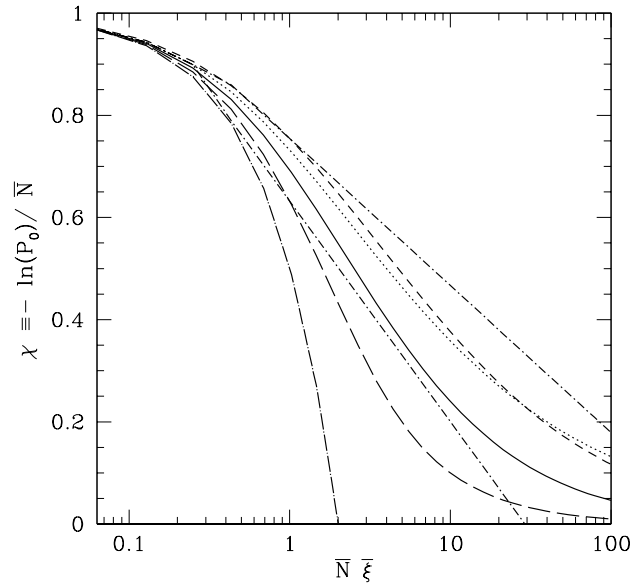}
\caption{Reduced void probability $\chi$ for different models 
  (left to right at $\chi\sim 0.4$): Gaussian (long-dashed-dotted,
  Eq.~12), minimal (long-dashed, Eq.~7), BBGKY (short-dashed-dotted,
  Eq.~10 with $Q=2/3$), negative binomial (continuous, Eq.~8), 
  thermodynamic (dotted, Eq.~9), lognormal (short-dashed) and BBGKY
  (short-dashed-dotted, Eq.~10 with $Q=1$). }
\label{models}
\end{figure}

In presenting our reduced VPF results, we follow the lead of Fry
(1986) and Fry et al. (1989) and compare with a number of model
scaling relations that differ in the way they fix the scaling
coefficients $S_p$. We give a brief description of these models here,
and refer the reader to the cited papers and references therein for
further details. In Fig.~\ref{models} we summarise the behaviour of each.

\subsubsection{Minimal model}

The first model is the so-called \emph{minimal cluster model}, the
motivation of which is to consider a clumpy galaxy distribution of
clusters, the cluster distribution in space itself being Poisson
with a Poisson galaxy occupancy. This is reminiscent of the halo model
(e.g. Cooray \& Sheth 2002) but with a Poisson halo/cluster profile.
Evaluating the set of $S_p$'s from the distribution function generated
by this model leads to a functional form for $\chi$ of
\bea{minimal}
\chi &=& (1-e^{-\bar{N} \bar{\xi}})/\bar{N} \bar{\xi}\ \ \ \rm{(minimal),} \\ 
\nonumber
S_p &=& 1   ~~~({\rm Skewness:}~ S_3=1)~.
\eea
Fry (1986) speculated that this model represents a lower bound on the
allowable functions $\chi(\bar{N} \bar{\xi})$ in any consistent
hierarchical model.

\subsubsection{Negative binomial model}

The second model, commonly called the \emph{negative binomial model},
has been used in a number of fields with different physical
motivations (Klauder \& Sudarshan 1968, Carruthers \& Shin 1983,
Carruthers \& Minh 1983, Fry 1986, Elizalde \& Gazta\~naga 1992,
Gazta\~naga \& Yokoyama 1993).
After a set of $T$ independent trials with probability $q$ for
``success'' and \smash{$p=1-q$} for ``failure'', the probability of
having $S$ number of successes \emph{and} \smash{$F=T-S$} number of
failures is given by the binomial distribution: \smash{$P(S) =
(F+S)!/S!/F!  (1-q)^F q^S$}.  The negative binomial distribution
describes the probability for having $S$ number of successes
\emph{after} a fixed number $F$ of failures: \smash{$P(S) =
(F+S-1)!/S!/(F-1)!  (1-q)^F q^S$}. Note that in the binomial case what
is fixed is the total number of trials.

\begin{table*}
  \centering
  \footnotesize
  \caption[]{Properties of the combined 2dFGRS SGP and NGP volume limited
  catalogues (VLCs). Columns 1 and 2 give the faint and bright
  absolute magnitude limits that define the sample. Column 3 gives the
  median magnitude of the sample, computed using the Schechter
  function parameters quoted by Norberg et al. (2002).  Columns 4, 5
  and 6 give the number of galaxies, the mean number density and the
  mean inter-galaxy separation for each VLC, respectively.  Columns 7
  and 8 state the redshift boundaries of each sample for the nominal
  apparent magnitude limits of the survey; columns 9 and 10 give the
  corresponding comoving distances. Finally, column 11 gives the
  combined SGP and NGP volume.  All distances are comoving and are
  calculated assuming standard cosmological parameters
  ($\Omega_{m}=0.3$ and $\Omega_{\Lambda}=0.7$).}
  \begin{tabular}{cccrccrrrrc} 
    \hline \hline
    \multicolumn{2}{c}{Mag. range} & Median mag. &  
           \multicolumn{1}{c}{N$_{\rm G}$} &
           {$\rho_{ave}$} & {d$_{mean}$} & {z$_{min}$} & {z$_{max}$} &
           {D$_{min}$} & {D$_{max}$} & {Volume}\\
    \multicolumn{2}{c}{\tiny $M_{b_{\rm J}}-5\log_{10}h$} & 
           {\tiny $M_{b_{\rm J}}-5\log_{10}h$}& & 
           {\tiny $10^{-3}/h^{-3}$Mpc$^3$} & {\tiny $h^{-1}$Mpc} & & &
           {\tiny $h^{-1}$Mpc} & {\tiny $h^{-1}$Mpc} & 
           {\tiny $10^6h^{-3}$Mpc$^3$}\\
           \hline
    $-$18.0 & $-$19.0 & $-$18.44 & 23290 & 9.26 & 4.76 & 0.014 & 0.088 & 39.0 & 255.6 & 2.52\\
    $-$19.0 & $-$20.0 & $-$19.39 & 44931 & 5.64 & 5.62 & 0.021 & 0.130 & 61.1 & 375.6 & 7.97\\
    $-$20.0 & $-$21.0 & $-$20.28 & 33997 & 1.46 & 8.82 & 0.033 & 0.188 & 95.1 & 537.2 & 23.3\\
    $-$21.0 & $-$22.0 & $-$21.16 &  6895 & 0.110 & 20.9 & 0.050  & 0.266 & 146.4 & 747.9 & 62.8\\
   \hline \hline
  \end{tabular}
\end{table*}

We can identify a ``success'' as finding a galaxy in a cell, so that
\smash{$P_N=P(N=S)$} is the CPDF.  The fixed number of failures, $F$,
is assumed to be inversely proportional to $\xibar$ (the larger the
$\xibar$, the smaller the number of failures to count a galaxy in a
cell).  The probability for a failure $p$ is assumed to be
proportional to the product $\Nbar\xibar$ (because of clustering there
is an $\Nbar\xibar$ rms excess of galaxies within a cell with $\Nbar$
density: the larger this clumpiness the larger the probability to miss
galaxies in a random cell).  After fixing the proportionality
constants, this leads to \smash{$F=1/\xibar$} and
\smash{$p=\Nbar\xibar/(1+\Nbar\xibar)$} (for a different derivation
see Gazta\~naga \& Yokoyama 1993).  This model is a discrete version
of the gamma probability distribution (see Gazta\~naga, Fosalba \&
Elizalde 2000). The reduced VPF and cumulants in this case are:
\bea{Negative}
\chi &=& \ln (1+\bar{N} \bar{\xi})/\bar{N} \bar{\xi}\ \ \ \rm{(negative\
binomial),} \\ \nonumber
S_p &=& (p-1)!  ~~~({\rm Skewness:}~ S_3=2)~.
\eea

\subsubsection{Thermodynamic model}

The third model was first suggested by Saslaw and Hamilton (1984) and
arose from a thermodynamic theory of the properties of gravitational
clustering. The original model had a fixed degree of virialization
(temperature or density variance) for all cell sizes, but such
behaviour is inconsistent with observations. The model was later
extended (see e.g. Fry 1986) to include a different level of
virialization at each scale, to be identified with the variance
$\xibar$ as a function of scale. The results is:
\bea{Thermo}
\chi &=& [(1+2\bar{N} \bar{\xi})^{1/2}-1]/\bar{N} \bar{\xi}\ \ \
\rm{(thermodynamic),} \\ \nonumber
S_p &=& (2p-3)!! ~~~({\rm Skewness:}~ S_3=3)~,
\eea
where $(2p-3)!!=(2p-3).(2p-5).(2p-7)...$ and truncates at zero.

\subsubsection{Lognormal distribution}

The \emph{lognormal} distribution (e.g. Coles \& Jones 1991, Weinberg \&
Cole 1993), is often used as a phenomenological model for galaxy and
dark matter clustering. Although no analytic expression exists for the
reduced void probability, it can be estimated numerically (see
above references) and is found to behave similarly to the
thermodynamic model, as shown in Fig.~\ref{models} (note how the
dotted and the short-dashed lines overlap).  As in the thermodynamic
model, the lognormal distribution also has a large skewness: $S_3 = 3
+ \bar{\xi}$ (which exactly tends to the thermodynamical value $S_3
\rightarrow 3$ on large scales where $\bar{\xi} \rightarrow 0$). In
fact, it should be noted that the lognormal model is not truly
hierarchical, as it does not have constant moments $S_p$, but in
practice the variations have little effect on the reduced void
distribution.

\subsubsection{BBGKY model}

The BBGKY model of Fry (1984) provides a prescription for $\chi$ and $S_p$
as an asymptotic solution to the BBGKY kinetic equations:
\bea{BBGKY}
\chi &=& 1 - (\gamma + \ln 4Q \bar{N} \bar{\xi})/8Q\ \ \ \rm{(BBGKY),} \\ 
\nonumber
S_p &=&  (4Q)^{p-2} {p\over{2(p-1)}}~,
\eea
where \smash{$\gamma=0.57721...$} is Euler's constant.  This
asymptotic solution is only a good approximation for large values of
$\Nbar\xibar$.  When $\Nbar\xibar$ becomes small, for completeness we
simply match it to the nearest model. 

The skewness in the BBGKY model contains a free parameter,
\smash{$S_3= 3 Q$}, with the restriction that \smash{$Q>1/3$}.  Fry
(1984) used \smash{$Q \simeq 1$}, which was close to the then observed
\smash{$S_3 \simeq 3$} value measured from the 3-point function in
real space (inferred from projected maps).  Croton et al. (2004) and
Baugh et al. (2004) have since shown that $S_3$ is in fact closer to
\smash{$S_3=2$} in the 2dFGRS, corresponding to the case where
\smash{$Q= 2/3$}. Both possibilities are shown as short-dashed-dotted
lines in Fig.~\ref{models}, with the upper curve for \smash{$Q=1$} and
the lower curve for \smash{$Q=2/3$}.  Since we later show that neither
of these $Q$ values with the BBGKY model are able to match the data
very well, for the sake of clarity we omit the lower \smash{$Q=2/3$}
curve in subsequent figures.  The upper curve is retained in order to
demonstrate the range of possible $\chi$ values that a hierarchical
model may have.

\subsubsection{Poisson and Gaussian distributions}

In addition to the above models we also use the analytic expressions
of the reduced VPF for purely Poisson and Gaussian distributions.
Trivially, from Eq.~6 we see that
\begin{equation}
\chi=1\ \ \ \rm{(Poisson),}
\end{equation}
and
\bea{Gaussian}
\chi &=& 1-\frac{1}{2} \bar{N} \bar{\xi}\ \ \ \rm{(Gaussian),} \\
\nonumber
S_p &=& 0  ~~~({\rm Skewness:}~ S_3=0)~.
\eea
The later only makes sense for small values of $\Nx$, but note that
even when the underlying distribution is not Gaussian, the above
expression always gives a good approximation to the void probability
in the limit of small  $\bar{N} \bar{\xi}$.

\section{The Data Sets}

\subsection{The 2dFGRS Data Set}

In our analysis we use the completed 2dFGRS (Colless et~al.\
2003). The catalogue is sourced from a revised and extended version of
the APM galaxy catalogue (Maddox et al. 1990), and the targets are
galaxies with extinction-corrected magnitudes brighter than
b$_{J}$=19.45.  Our galaxy sample contains a total of 221,414 high
quality redshifts. The median depth of the full survey, to a nominal
magnitude limit of $b_{\rm J} \approx 19.45$, is $z \approx 0.11$. We
consider the two large contiguous survey regions, one in the south
Galactic pole (SGP) and one towards the north Galactic pole (NGP), and
restrict our attention to the parts of the survey with high redshift
completeness ($>70\%$). Full details of the 2dFGRS and the
construction and use of the mask quantifying the completeness of the
survey can be found in Colless et~al.\ (2001, 2003).

A model accounting for the change in galaxy magnitude due to
redshifting of the b$_J$-filter bandpass (k-correction) and galaxy
evolution (e-correction) was adopted following Norberg et al. (2002):
\begin{equation}
k(z) + e(z) = \frac{z + 6z^2}{1 + 20z^3}~.
\end{equation}
This model gives the mean k+e-correction over the mix of different
spectral types observed in the 2dFGRS sample, and was shown by 
Norberg et al. to accurately account for such observational effects when
estimating 2dFGRS galaxy absolute magnitudes.

\subsection{Volume Limited Catalogues}

The 2dFGRS galaxy catalogue is \emph{magnitude-limited}, meaning the
survey is constructed by observing galaxies brighter than the fixed
apparent magnitude limit of b$_{J}$=19.45.  A magnitude-limited galaxy
catalogue is not uniform in space, since intrinsically fainter objects
may be missed even if they are relatively nearby, while the most
luminous galaxies will be seen out to large distances.  This
non-uniformity of the magnitude-limited catalogue must be dealt with
for a correct statistical analysis, and the simplest way to do this
with a catalogue the size of the 2dFGRS is by constructing a volume
limited catalogue (VLC) from the magnitude-limited sample.

Volume limited catalogues are defined by choosing minimum and maximum
\emph{absolute} magnitude limits.  These limits, along with the
intrinsic apparent magnitude limits of the survey, define minimum and
maximum redshift boundaries via standard luminosity--distance
relations (Peebles 1980). The VLC is built by selecting galaxies whose
redshift lies within the minimum and maximum boundaries just
determined, and whose absolute magnitude lies within the specified
absolute magnitude limits.  Such galaxies can be displaced to any
redshift within the VLC volume and still remain within the bright and
faint apparent magnitude limits of the magnitude limited survey.
Table~1 presents the properties of the combined NGP and SGP
volume limited catalogues used in this paper.

\section{Measuring the Galaxy Distribution}

To measure the void probability function we use the method of
counts-in-cells.  The survey volume is uniformly sampled with a large
number ($2.5\times 10^7$) of randomly placed spheres of fixed radius
$R$, and we record the number of times a sphere contains exactly $N$
galaxies.  Our choice of massive oversampling ensures a high level of
statistical accuracy in the calculation (Szapudi 1998).  The CPDF can
then be found as the probability of finding exactly $N$ galaxies in a
randomly placed sphere:
\begin{equation}
P_N(R) = \frac{N_N}{N_T}~,
\end{equation}
where $N_{N}$ is the number of spheres that contain exactly $N$
galaxies out of the total number of spheres thrown down, $N_{T}$.  By
definition, the void probability function is the probability of
finding an empty sphere:
\begin{equation}
P_0(R) = \frac{N_0}{N_T}~.
\end{equation}
The mean number of galaxies expected inside a sphere of radius $R$ is
readily calculated from
\begin{equation}
\bar{N}(R)= \sum N P_N(R)~,
\end{equation}
and this estimation of $\bar{N}$ for each individual VLC is found to
be independent of scale and indistinguishable from that determined
from the known mean galaxy density.  The volume averaged $2$-point
correlation function, $\bar{\xi_2}$, is found directly from the second
moment of the CPDF:
\begin{equation}
\bar{\xi_2}(R) = \frac{\langle (N-\bar{N})^2 \rangle - \bar{N}(R) }{\bar{N}(R)^2}~.
\end{equation}
We have also carried out an independent counts-in-cells analysis by
placing the spheres at the positions of a regular spatial lattice that
homogeneously oversamples the survey area. The results are insensitive
to these details.

The 2dFGRS has an inherent spectroscopic galaxy incompleteness which
will change the results of any void analysis (Colless et al. 2001).
In addition, due to the irregular geometry of the survey boundaries it
is difficult to guarantee that every sphere will be completely
contained within the regions we wish to measure.  Since the CPDF is
sensitive to such effects we adopt a technique which accurately
accounts for such deficiencies.  This method is explained and tested
in Appendix A (see also Croton et al. 2004).

\subsection{Error Estimation}

We estimate the error on our void statistics using the set of 22 mock
2dFGRS surveys described by Norberg et al. (2002; see also Cole et
al. 1998).  These mock catalogues have the same radial and angular
selection function as the 2dFGRS and have been convolved with the
completeness mask of the survey.  The mock catalogues are drawn from
the Virgo Consortium's $\Lambda$CDM Hubble Volume simulation and thus
include sample variance due to large scale structure (see Evrard et
al. 2002 for a description of the Hubble Volume simulation).  The $1
\sigma$ errors we quote correspond to the \emph{rms} scatter over the
ensemble of mocks (see Norberg et al. 2001).  We have compared this
estimate with an internal estimate using a jack knife technique
(Zehavi et al. 2002). In the jack knife approach, the 
survey is split into subsamples. The error is then the scatter between
the measurements when each subsample is omitted in turn from the
analysis.  The jack knife gives comparable errors to the mock ensemble
for the VPF measurement.

\section{Results}

We begin with Fig.~2, where we plot the reduced void probability
function, $\chi$, individually as a function of both the mean galaxy
number, $\bar{N}$, and the variance, $\bar{\xi}$, in the top and
bottom panels respectively.  The physical scale given on each top axis
corresponds to values for the \smash{$-20>M_{b_{\rm J}}-5\log_{10}>-21$} VLC
only, and is included for reference (for VLCs of different mean
density the scale at which a given $\bar{N}$ or $\bar{\xi}$ will occur
will be different).
Note that for VLCs fainter than our reference this scale shifts to the
right in the top panel and to the left in the bottom panel.  The
converse is true when considering brighter galaxies than the
reference.

The main feature of this figure is that neither $\bar{N}$ nor
$\bar{\xi}$ individually show hierarchical scaling when plotted
against $\chi$.  Note that smaller values of $\chi$ correspond to
larger deviations from a Poisson distribution.  Brighter galaxy
samples show behaviour which is closer to that of the Poisson
distribution for any given value of $\bar{N}$ or $\bar{\xi}$, however
this merely reflects the fact that the brightest VLCs are also the
sparsest (Table~1).

We now test for hierarchical scaling in the 2dFGRS, as outlined in
Section~2.2.  In Fig.~\ref{fig:chi} we plot the reduced void
probability function, $\chi$, as a function of the scaling variable
$\bar{N} \bar{\xi}$.  In this way we eliminate the dependence of the
void probability on the variance and mean density.  This figure shows
VLCs ranging in absolute magnitude from $-18$ to $-22$.  If a scaling
between correlation functions of different orders exits we expect to
see all points for each catalogue fall onto the same line.  Again we
provide a reference scale on the top axis, given for the
\smash{$-20>M_{b_{\rm J}}-5\log_{10}>-21$} VLC, and note that for
fainter galaxy samples this scale shifts to the right and conversely
for brighter samples.  Over-plotted are the scaling models previously
discussed in Section~2.3: (bottom to top) the Gaussian (Eq.~12),
minimal (Eq.~7), negative binomial (Eq.~8), thermodynamic
(Eq.~9), lognormal, and BBGKY (Eq.~10, \smash{$Q=1$}) models
respectively. 

\begin{figure}
  \plotone{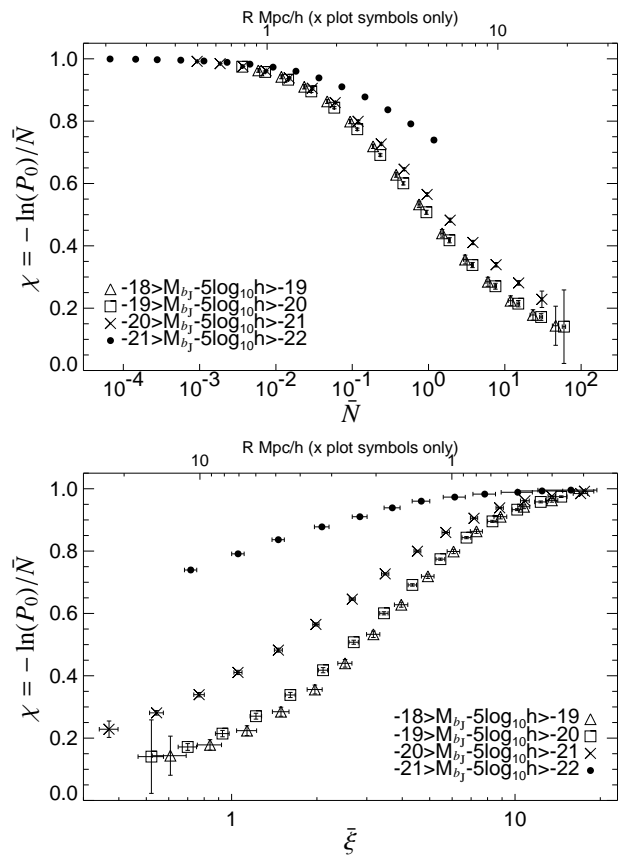}
  \caption{The 2dFGRS reduced VPF, $\chi=-\ln P_0/\bar{N}$, as a function of
    (top) the mean galaxy number, $\bar{N}$, and (bottom) the variance
    of the distribution, $\bar{\xi}$, as measured for volume limited
    catalogues in varying luminosity bins (Table~1). Smaller values of
    $\chi$ imply larger deviations from a Poisson distribution.  The
    reference scale given on the top axis is for the $-20>M_{b_{\rm
        J}}-5\log_{10}>-21$ VLC only (each $\bar{N}$ and $\bar{\xi}$ value
    individually correspond to different scales for each VLC).  Notice
    that neither variable displays hierarchical scaling when plotted
    individually against $\chi$.
    }
\end{figure}

\begin{figure*}
\plotfull{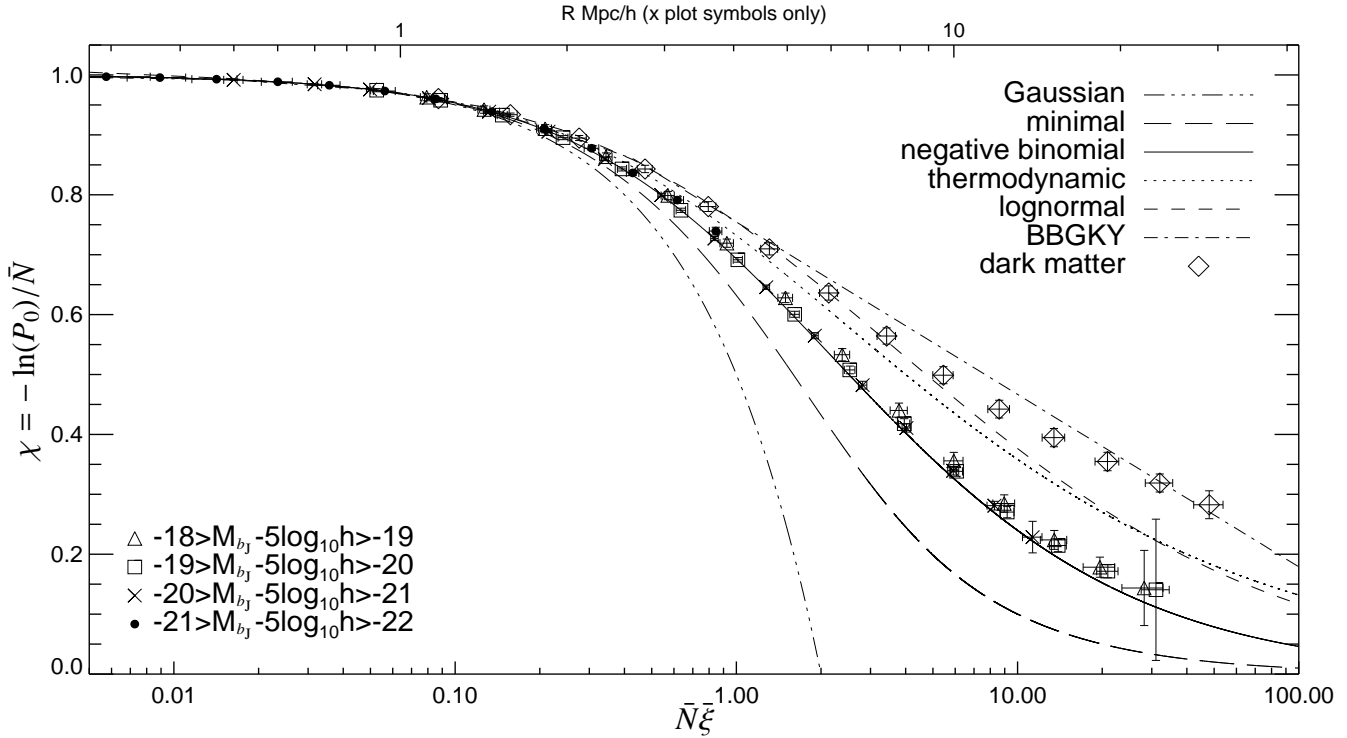}
\caption{The reduced VPF, $\chi=-\ln P_0/\bar{N}$, as a function of the
  scaling variable $\bar{N} \bar{\xi}$ for the four 2dFGRS galaxy VLCs from
  Table~1. The dark matter reduced VPF, as measured from the $\Lambda$CDM Hubble
  Volume simulation, is shown as large diamonds.  In all cases, smaller
  values of $\chi$ imply larger deviations from a Poisson
  distribution.  The reference scale given on the top axis is for the
  \smash{$-20>M_{b_{\rm J}}-5\log_{10}>-21$} VLC only (each $\bar{N}
  \bar{\xi}$ value corresponds to a different scale for each VLC).  If
  hierarchical scaling is present in the galaxy distribution all
  points should collapse onto a single line, which is clearly
  seen. The six curves represent the hierarchical models discussed in
  Section 2.3 (Eq.~7 to 12).
  \label{fig:chi}
  } 
\end{figure*}

\begin{figure}
\plotone{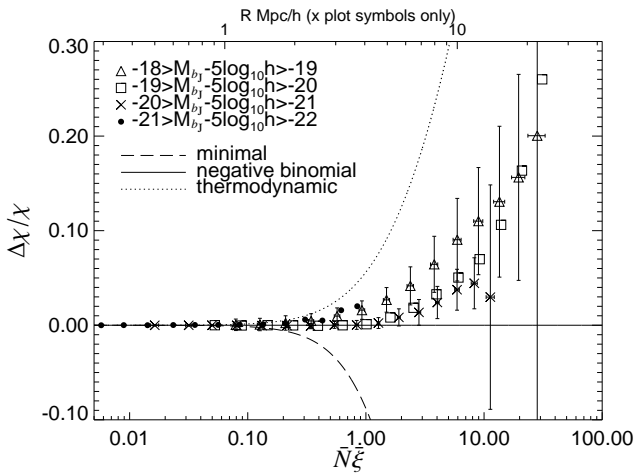}
\caption{The fractional difference between the negative binomial model
  and the 2dFGRS, thermodynamic, and minimal reduced VPFs.  The
  reference scale given on the top axis is for the
  \smash{$-20>M_{b_{\rm J}}-5\log_{10}>-21$} VLC only (each $\bar{N}
  \bar{\xi}$ value corresponds to a different scale for each VLC).
  Some error bars have been omitted for clarity.  All 2dFGRS results are
  consistent with the negative binomial model at the $2\sigma$ level.}
\end{figure}

Fig.~\ref{fig:chi} demonstrates the clear signature of hierarchical
scaling in the clustering moments of the 2dFGRS.  All points are seen
to follow a tight path (within the error bars) out to values of
$\bar{N} \bar{\xi} \sim 30$, and sit close to the negative binomial
model prediction along this entire range.  Such values encompass
galaxy clustering from the deeply non-linear to the linear regime,
revealing hierarchical scaling out to scales of $\sim 20h^{-1}$Mpc or
more. 

For comparison, in Fig.~3 we also present the dark matter reduced VPF
measured from the $\Lambda$CDM Hubble volume simulation (particle mass
$2.3 \times 10^{12}h^{-1}M_{\odot}$) (Evrard et al. 2002).  We
independently analyse $100$ randomly placed cubes of side length
$200h^{-1}$Mpc (approximately equal in volume to our $M^*$ galaxy
volume limited sample), from which the rms is then plotted.  In
contrast to the 2dFGRS galaxies, the dark matter follows a lognormal
distribution out to values of $\bar{N} \bar{\xi} \sim 6$ (a scale of
approximately $R \sim 4h^{-1}$Mpc in the simulation), but then
deviates strongly on larger scales (the last point plotted corresponds
to $R = 10h^{-1}$Mpc in the simulation).
 
To highlight the differences between the 2dFGRS galaxy reduced VPF and
the negative binomial prediction, in Fig.~4 we show the fractional
difference between the two.  Also included are the ``bounding'' models
closest to the negative binomial: the minimal and thermodynamic
models.  All 2dFGRS points plotted are consistent with the negative binomial
model at the $2\sigma$ level.  At larger values of $\bar{N} \bar{\xi}$
we find some small departures from the negative binomial model, and it
is interesting to note that these deviations appear the greatest for
the faintest VLC.
This could be explained by the weak dependence of $S_p$ on galaxy
luminosity found by Croton et al. (2004), where fainter samples
typically had larger $S_p$ values than brighter samples (albeit with
large error bars).  The effect of such an increase in the hierarchical
picture would result in a value of $\chi$ closer to unity (Eq.~6).

An important feature of Fig.~\ref{fig:chi} is the inconsistency of the
reduced void probability function with a Gaussian distribution across
all scales considered (up to approximately $30h^{-1}$Mpc).  On large
scales where the galaxy correlation functions become too small to
measure independently, the value of $\bar{N}$ is found to increase
faster than $\bar{\xi}$ decreases, and thus $\chi$ is still affected
strongly by higher-order correlations.  It is clear that even in the
quasi-linear regime, where one would expect galaxy clustering to be
very simple, higher-order correlations still play a significant role
in the make-up of the large scale distribution.

To evaluate the robustness of the results seen in Fig.~\ref{fig:chi}
we apply two tests to illustrate the degree of confidence we should
have in believing the existence of hierarchical scaling in the 2dFGRS.
Firstly, one of the most valuable features of the 2dFGRS is that we
have available data from two totally independent regions on the sky,
the SGP and NGP.  So far we have been calculating our void statistics
from the combined volume of the two, but it is useful to check that
the scaling properties still exist in the two regions independently.
This we do in the top panel of Fig.5, where the large symbols
represent the SGP and small symbols the NGP.  It is immediately clear
that galaxies from both the SGP and NPG regions independently obey
hierarchical scaling and reproduce the negative binomial results
discussed previously to good accuracy.

Secondly, we test the scaling properties seen in Fig.~\ref{fig:chi} by
calculating the reduced VPF for randomly diluted samples of galaxies.
Such dilutions leave the $2$-point correlation function unchanged, and
within the hierarchical paradigm the scaling exhibited in
Fig.~\ref{fig:chi} should also remain unchanged.  This test is shown in
the bottom panel of Fig.~5, where we have diluted each of the VLCs
used in Fig.  ~\ref{fig:chi} by factors of $0.5$ (large symbols) and
$0.25$ (small symbols).  We again see that the trend for hierarchical
scaling exists and follows the negative binomial model, consistent
with our previous conclusions.

\begin{figure}
\plotone{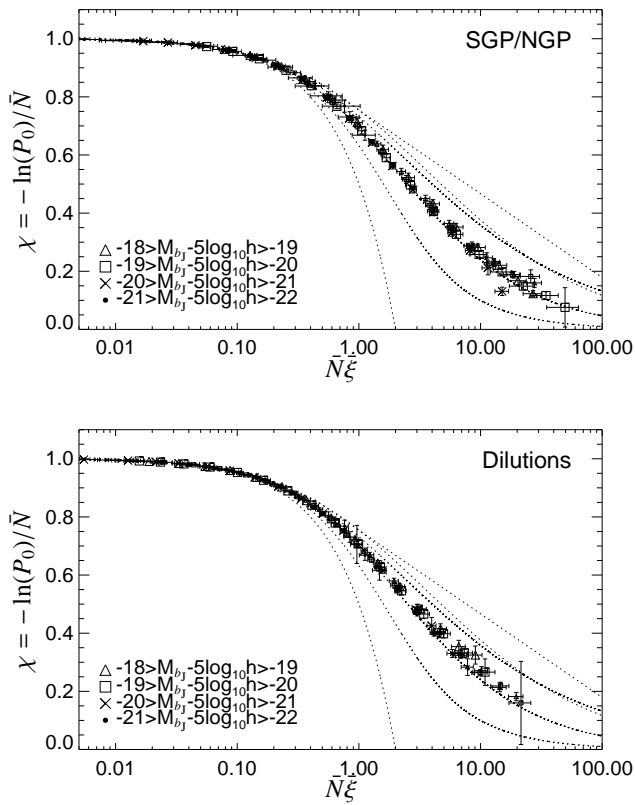}
\caption{Two tests of the scaling properties seen in Fig.~\ref{fig:chi}
 using the reduced VPF, $\chi=-\ln P_0/\bar{N}$, as a function of the
  scaling variable $\bar{N} \bar{\xi}$. (top) Independent SGP and NGP
  VLCs show identical scaling to that seen in Fig.~\ref{fig:chi}.
  Here the large symbols represent the SGP result, and the small
  represent the NGP result.  (bottom) The same combined VLCs as in
  Fig.~\ref{fig:chi}, but now diluted by factors of 0.5 (large
  symbols) and 0.25 (small symbols). If hierarchical scaling exists in
  the galaxy distribution, dilution should make little difference
  to the results found in Fig.~\ref{fig:chi}. For both panels, the
  dotted curves represent the same six models plotted in Fig.~3 and
  discussed in Section~2.3.  Some error bars have been omitted for
  clarity. 
  }
\end{figure}

\section{Discussion}

The 2dFGRS represents an enormous improvement in volume and number of
galaxies over previous surveys, such as the CfA or the SSRS
samples. Here we measure the galaxy distribution over both a wider
range in variance ($\bar{\xi} \sim 0.3 - 20$) and mean galaxy number
($\bar{N} \sim 10^{-4} -
10^2$).  The impact on the VPF can be seem by comparing Fig.~3 above
to Fig.~7 in Gazta\~naga \& Yokoyama (1993), where the CfA and SSRS
data can not discriminate between the negative binomial and the
thermodynamical models. As shown here in Fig.~3 and 4, although the
agreement is not always perfect, the negative binomial does much
better, by far, than any of the other models considered in the
literature.  This includes the lognormal distribution, which is close
to the thermodynamical model (Fig.~1) and is widely used as a
phenomenological clustering model.  These results are valid
independently in the NGP and SGP regions of the survey, and do not
change when we randomly dilute the galaxy samples (Fig.~5).  
The lognormal distribution does, however, appear to be a good
representation for the distribution of dark matter on smaller scales
(less than $\sim 4h^{-1}$Mpc), although not at larger scales.
The differences between the galaxy and dark matter reduced VPFs can be
understood by noting the differences between their higher-order
volume-averaged correlation functions, as shown by Baugh et al. (2004).

The 2dFGRS reduced void probability function appears to behave
differently from the one presented by Vogeley et al. (1994) for the
CfA-1 and CfA-2 samples, which show more scatter with magnitude and
values well above the negative binomial model (compare their Fig.~4 to
our Fig.~\ref{fig:chi}). Here we do not observe any significant
departure from the scaling models on scales larger than $R\sim
8.5h^{-1}$Mpc as they had previously found.  In contrast, our results
indicate hierarchical scaling exists in the galaxy distribution out to
scales of at least $R\sim 20h^{-1}$Mpc.

Although some heuristic derivations exist for the negative binomial
distribution (see section 2.3), we have not found a satisfactory
physical explanation for the very good performance of this model. The
value of the skewness for the negative binomial model, $S_3=2$, is
quite close to the direct measurement in the 2dFGRS: $S_3 = 1.86-2.03$
(Baugh et al. 2004). Other phenomenological models, such as the
thermodynamical or the lognormal distribution, have larger values for
the skewness ($S_3 \simeq 3$). A similar trend was found by Baugh et
al. for the higher order coefficients $S_4$, $S_5$, and $S_6$.  In
this respect it is not totally surprising that the negative binomial
does better.  The one freedom the reduced VPF has is in the value of
the scaling coefficients which appear in the sum in Eq.~6.  If these
coefficients are found to match that predicted by a particular
hierarchical scaling model, then one would expect their reduced
VPFs to look similar.

Perturbation theory with Gaussian initial conditions predict values
for the $S_p$'s that are universal and only depend on the local
spectral index.  They are therefore a known function of scale. Such
scale dependence, however, breaks the hierarchy in Eq.~3, and
therefore the universality of the scaling in Eq.~6.  On the other
hand, redshift space distortions and biasing tend to wash away this
scale dependence (see e.g.  Fig.~49 in Bernardeau et al. 2002), an
argument which has been used to explain the good performance of the
scaling hierarchy.  But, as shown by Baugh et al.  (2004) and Croton
et al. (2004), the measured values of the $S_p$'s do not seem to match
the expectations in either dark matter models or mock galaxy surveys
(both in redshift space).  The reasons for this, and a more physically
motivated interpretation of the negative binomial model, will provide
important constraints to be matched by models of galaxy formation.

\section*{Acknowledgements}

We wish to thank everyone involved with the 2dF instrument and
Anglo-Australian telescope for their sustained effort over many years
in producing the 2dFGRS.  Many thanks also go to Saleem Zaroubi, Simon
White, and Andrew Benson.  DC acknowledges the financial support of
the Mount Stromlo Bok Honours Scholarship, the Australian National
University Coombs Honours Scholarship, and the International Max
Planck Research School in Astrophysics Ph.D. fellowship.  EG
acknowledges support from the Spanish Ministerio de Ciencia y
Tecnologia, project AYA2002-00850 and EC FEDER funding.
CMB is supported by a Royal Society University Research Fellowship.  
PN acknowledges receipt of a Zwicky Fellowship.

\section*{APPENDIX A: CORRECTING FOR INCOMPLETENESS IN THE 2dFGRS} 

The 2dFGRS is spectroscopically incomplete to a small degree resulting
in missed galaxies (see Colless et al. 2001), and some spheres used in
our counts-in-cells analysis may straddle the survey boundaries or
holes resulting in missed volume.  Such influences will induce an
artifical ``voidness'' that will be picked up by our VPF measurements,
and any analysis that neglects these effects will tend to over predict
the VPF.  Thus it is desirable to devise a method with which one can
confidently correct for such incompleteness.  This is not a trivial
exercise, since weighting schemes that work with other statistics
(e.g. Efstathiou et al. 1990) cannot necessarily be applied here, as
the VPF will remain uncorrected (how does one weight no galaxies?).
Such techniques will lead to an under-estimation of the mean density
of galaxies and an over-estimation of the influence of the voids.
Ideally, we need to ensure that any correction faithfully reproduces
the full CPDF of the complete distribution for all orders of galaxy
clustering.

\begin{figure}
\plotone{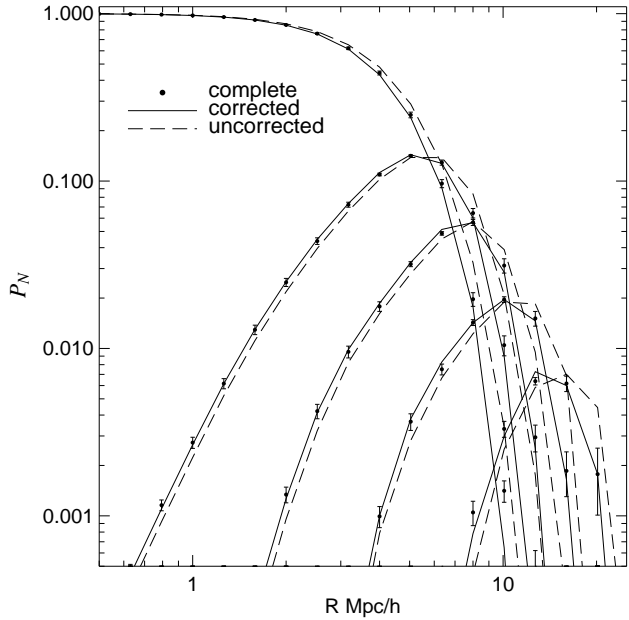}
\caption{ Correcting for incompleteness in the 2dFGRS. The CPDF,
  $P_N$, for a Hubble Volume 2dFGRS mock VLC in the magnitude range
$-19>M_{b_{\rm J}}-5\log_{10}>-20$: left to right $N=0$ (the VPF), $2$, $6$,
$20$ and $70$. The points with errors represent the complete mock, the
solid line is the corrected incomplete mock and the long-dashed line
is the uncorrected incomplete mock.  Note the uncorrected incomplete
mock always lies outside the error bars.}  
\end{figure}

To resolve these problems we have adopted the following method.  When
a satisfactory sphere location is found in the 2dFGRS wedge we project
the sphere onto the sky and estimate, using the survey masks (Colless
et al. 2001), the average completeness $f$ within the sphere.  Due to
the incompleteness effects described above we typically will have
\smash{$f<1$}.  Instead of viewing this incompleteness as missed
galaxies, we instead consider it as \emph{missed volume}, and to
compensate we scale the radius of the sphere according to \smash{$R' =
R / f^{\frac{1}{3}}$}.  This new radius gives an effective sphere
volume \emph{with incompleteness} equal to that of a $100\%$ complete
sphere with the original radius.  Galaxies are counted within the new
radius $R'$, but contribute their counts to the scale $R$.  Each
sphere we place is individually scaled in this way according to its
local incompleteness, as given by the masks.  We note that due to our
chosen acceptable minimum incompleteness of $0.7$ spheres are never
scaled beyond the radius bin $R$ under consideration.  Thus each
correction applies only to the value of the VPF at each radius point
plotted.

We have tested the robustness of our method by comparing measurements
of the CPDF using a fully sampled, complete Hubble Volume 2dFGRS mock
VLC (Norberg et al. 2002) with those from the same mock but which have
been made artifically incomplete using the survey masks
(spectroscopically, and including irregular boundaries and holes) and
then corrected.  In Fig.~6 we show the results for $P_N$ vs. radius,
where $N=0$ (the VPF), $2$, $6$, $20$ and $70$ (note other $N$'s are
omitted for clarity, but all behave similarly over the scales where
the VPF is of interest to us).  Here the points with error bars are
the complete $P_N$'s, the solid lines are the equivalent corrected
incomplete $P_N$'s, and the dashed lines represent the uncorrected
incomplete $P_N$'s.  As can be seen, the complete points and corrected
lines are fully consistent, whereas the uncorrected values almost
always lie off the complete points and well outside their error bars
(note the steepness of each curve which is plotted on a log
scale). The $P_0$ curve in particular demonstrates that such
incompleteness effects must be accounted for to obtain correct void
measurements; simply building volume limited catalogues is not enough
and will lead to an over-prediction of the scale and frequency of
voids in the survey.  Our method can be applied to any counts-in-cells
analysis where incompleteness in the galaxy distribution is present.

\label{lastpage}

\end{document}